\documentclass[twocolumn,showpacs,epsfig,pre]{revtex4}

\usepackage{graphicx}
\usepackage{amssymb}
\usepackage{color}

\newcommand{\figsize}{0.48}
\newcommand{\figsizetwo}{0.35}

\begin{document}

\draft
\title{Quasiattractors in coupled maps and coupled dielectric cavities}
\author{Jung-Wan Ryu$^1$, Martina Hentschel$^2$, and Sang Wook Kim$^3$}
\email{swkim0412@pusan.ac.kr}
\affiliation{$^1$Department of Physics, Pusan National University, Busan 609-735, South Korea\\
$^2$Max-Planck-Institut f\"ur Physik komplexer Systeme, N\"othnitzer Str. 38, D-01187 Dresden, Germany\\
$^3$Department of Physics Education, Pusan National University, Busan 609-735, South Korea}
\date{\today}

\begin{abstract}
We study the origin of attracting phenomena in the ray dynamics of coupled optical microcavities. To this end we investigate a combined map that is composed of standard and linear map, and a selection rule that defines when which map has to be used. We find that this system shows attracting dynamics, leading exactly to a quasiattractor, due to collapse of phase space.
For coupled dielectric disks, we derive the corresponding mapping based on a ray model with deterministic selection rule and study the quasiattractor obtained from it.
We also discuss a generalized Poincar\'e surface of section at dielectric interfaces.
\end{abstract}
\maketitle
\narrowtext

\section{Introduction}

An attractor in classical dynamics is, colloquially speaking, a (invariant) set of points to which all trajectories starting in its neighborhood (more precisely, its basin of attraction) converge. More precisely, an attractor can be defined as a closed set
$A$ with the following properties \cite{Ber84,Str94,com01}. First, $A$ is an invariant set: any trajectory $\mathbf{x}(t)$ that starts in $A$ stays in $A$
for all time. Secondly, $A$ attracts an open set of initial conditions: there is an open set $U$ containing $A$ such that if
$\mathbf{x}(0) \in U$, then the distance from $\mathbf{x}(t)$ to $A$ tends to zero as $t \rightarrow \infty$. This means that $A$
attracts all trajectories that start sufficiently close to it. The largest such $U$ is called the basin of attraction of $A$ \cite{remark_A-is-minimal}.

The fact that volumes have to be conserved in conservative systems implies immediately that they display no attracting regions
in phase space \cite{Lic92,Sto99}.
However, a quasidissipative property has been reported in a combined map, namely a piecewise smooth area-preserving map which models an electronic relaxation oscillator with over-voltage protection \cite{Wan01}.
The quasidissipative property eventually converts initial sets into attracting sets, which was correspondingly referred to as the formation of a quasiattractor.
As a result, this system behaves partly dissipative (outside the quasiattractor, i.e. before trajectories have reached the quasiattractor) and partly conservative (inside the quasiattractor).
To the best of our knowledge, the existence of quasiattractors has been observed so far only in the piecewise smooth area-preserving map. In this paper we show that coupled dielectric cavities are another system class with this property. 

Each optical interface is characterized by the splitting of an incoming ray into transmitted and reflected rays which is repeated at each reflection point. For optical systems consisting of more than one dielectric building block such as the combination of two disks, the ray splitting dynamics is especially complicated. 
A ray model with deterministic selection rule (RMDS) was proposed \cite{Ryu10} to effectively describe the resulting dynamics. Its justification is underlined by a nice agreement with wave calculations \cite{Ryu10}.

A noticeable and characteristic property of ray dynamics in the RMDS is that all initial rays eventually arrive in a certain region of phase space - namely an island structure \cite{Ryu10}. In other words, attractors occur. Their existence is, on the one hand, a surprise because the underlying ray dynamics is Hamiltonian. On the other hand, wave solutions for the individual resonant modes are precisely localized on these emerging (quasiattractor) structures, and ray-wave correspondence is fully established. We shall see below in Sec. \ref{sec_III} that it is the special structure of the selection rule, namely the coupling of different maps, that induces the quasiattracting features. 

In this paper, we study the characteristics of quasiattractors emerging from simple coupled maps in order to gain a heuristic understanding of attracting phenomena.
To this end, we introduce in Sec. \ref{sec_II} a toy model - a combination of standard and linear mappings - that clearly demonstrates the existence of the quasiattracting phenomenon and its origin. In Sec. \ref{sec_III} we apply the insight gained to the optical system consisting of two dielectric disks, and explicitly show the mapping rules and their relation to the quasiattractor. We use the dependence of the selection rule on the refractive index of the disks in order to illustrate the parametric dependence of the emerging structures. 
Finally, we summarize our results in Sec. \ref{sec_IV}.

\section{Origin of quasiattracting phenomena}
\label{sec_II}

\subsection{Case study: Combined standard-linear mapping}

\begin{figure}
\begin{center}
\includegraphics[width=\figsize\textwidth]{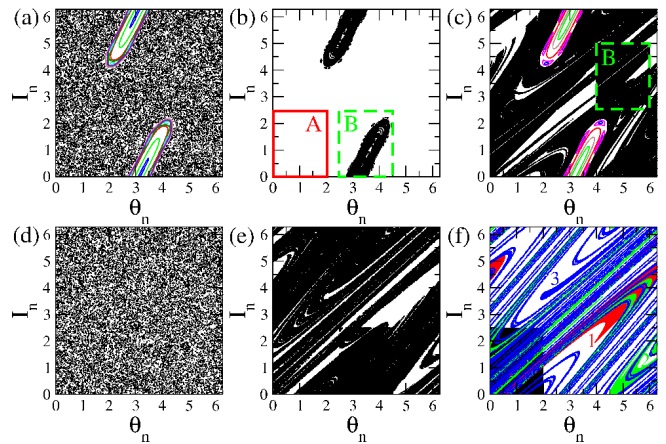}
\caption{(color online) PSOSs of standard mapping for (a) $K=4.0$ and (d) $K=9.0$.
Different color sets represent those from different initial points.
The PSOS of the combined standard-linear map; (b, c) for $K=4.0$ and (e, f) for $K=9.0$.
The region A is denoted as the red solid box in (b).
The region B, denoted as the dashed green box, is chosen as (b, e) $(\alpha,\beta) = (2.5, 0.0)$ and (c) $(\alpha,\beta) = (4.0, 2.5)$. 
(f) Forward iterated sets starting from the region A using only the standard map. Red, green, and blue sets represent the first, second, and third iterated sets, respectively.
}
\label{fig1}
\end{center}
\end{figure}

In order to understand the appearance of quasiattractors, we introduce a map that combines the well-known standard and linear maps as an example of a piecewise smooth area-preserving map discussed in Refs.~\cite{Wan01} as origin of quasidissipative behaviour. 
The standard map is, physically speaking, a discrete-time analogue of the equation of the vertical pendulum and a prototype two-dimensional map commonly used in the study of various nonlinear phenomena of conservative systems \cite{Chi79,Gre79}.
It is given by
\begin{eqnarray}
\label{eq1}
\theta_{n+1} &=& \theta_n + I_{n+1} ~\mathrm{(mod ~2\pi)},\\\nonumber
I_{n+1} &=& I_n + K \sin{\theta_n} ~\mathrm{(mod ~2\pi)}, \qquad \textrm{for} ~ (\theta_n,I_n) \notin \mathrm{A}.
\end{eqnarray}
where $\theta$ is the angle of rotation, $I$ is the conjugate momentum,
$K$ is a positive parameter which determines the dynamics of the map, and $\mathrm{A}$ is a subset of phase space introduced in connection with the selection rule. 
Equation~(\ref{eq1}) is the usual standard map if we exclude the selection rule, i.e., apply the map to all $(\theta_n,I_n)$.
For $K=0$, the momentum is constant and the angle increases linearly.
As $K$ increases, the phase space becomes increasingly chaotic as can be seen in the Poincar\'e surface of sections (PSOSs) in Fig.~\ref{fig1}.
For $K=4.0$, cf.~Fig.~\ref{fig1} (a), one (split) island of stability remains, whereas 
all islands have disappeared for $K=9.0$, see Fig.~\ref{fig1} (d). 

In addition to the standard map we introduce a linear map with parameters $\alpha, \beta$ that has to be determined when $(\theta_n,I_n) \in \mathrm{A}$,
\begin{eqnarray}
\label{eq2}
\theta_{n+1} &=& \theta_n + \alpha ~\mathrm{(mod ~2\pi)}, \\\nonumber
I_{n+1} &=& I_n + \beta ~\mathrm{(mod ~2\pi)}, \qquad \textrm{for} ~ (\theta_n,I_n) \in \mathrm{A}.
\end{eqnarray}

\subsection{Origin of the quasiattractor}

Each dynamics of standard maps and linear maps have been well understood for many years,  
but the dynamics of the combined map is not the case.
The region $0.0<\theta_n<2.0$ and $0.0<I_n<2.5$ is chosen for A, indicated by the red box in Fig.~\ref{fig1} (b).
With $(\alpha,\beta) = (2.5,0.0)$
the region A is mapped into the region B represented by the dashed box in Fig.~\ref{fig1}(b) via the linear map (2).
After some transient time, for $K=4.0$ the initial distribution uniformly prepared over the whole phase space converges into the island structure as shown in Fig.~\ref{fig1}(b).
This behavior can be easily understood as follows.
The points starting from the regions outside the converging islands eventually reach the regions A due to ergodicity of chaotic dynamics.
Some points of A are mapped into the islands according to the linear mapping of (2) since the region B contains the islands.
Once the points are put on the converging islands, they cannot escape from the islands since it forms an invariant set.
If they are not put on the islands, they chaotically wander again until they are mapped into the islands.
Therefore, the whole phase space converges to the islands as time goes on so that the islands form a quasiattractor \cite{Wan01}.

The structure of the (quasi)attractor depends on where the region A and B are located.
When B is chosen as the dashed box shown in Fig.1~(c) with $(\alpha, \beta) = (4.0 ,2.5)$,
the quasiattractor consists of the chaotic components as well as the islands.
The chaotic part resembles a usual strange attractor of dissipative systems \cite{Eck85}. Its structure will be discussed in detail below.

For $K=9.0$, no stable islands exist in the standard map.
After some transient time, a fractal quasiattractor emerges for $(\alpha, \beta) =  (2.5, 0.0)$ as shown in Fig.~\ref{fig1}(e).
In order to understand the detailed structure of the fractal quasiattractor, let us consider the evolution of the region A only by the standard map (1).
The first, second, and third iterations of the region A are denoted as red, green, and blue areas in Fig.~\ref{fig1}(f), respectively,
which shows a typical stretching and folding structure of chaotic dynamics.
Due to the linear map (2), the region A cannot follow such an evolution but is compulsorily mapped into B by (2).
It means that the final attracting structure is likely to lack the areas generated by chaotic evolution of the region A.
In fact, the empty area of Fig.~\ref{fig1}(e) looks quite similar to the collection of the first, the second, and the third iterations
of the region A shown in Fig.~\ref{fig1}(f). Note that this argument is only true for a short time evolution.
However, the dissipative nature of our system guarantees that the short time dynamics is enough to understand its main feature.
This mechanism can also be applied to Fig.~\ref{fig1}(c) although the island part is dealt with separately.

The fractal quasiattractors shown in Fig.~1(c) and (e) form an invariant set with a non-zero measure, which is called a fat fractal.
As a parameter such as the location of the region B varies,
the fractal quasiattractor loses its stability and at the same time the island suddenly becomes an attractor.
This is a typical characteristic of the so-called crisis bifurcation \cite{Wan02, He04}.
It is known that forming a quasiattractor the phase space decreases on average linearly in time rather than exponentially \cite{Jia04}.

\section{Quasiattractor in coupled dielectric cavities}
\label{sec_III}

The coupled dielectric cavities, schematically shown in Fig.~2, have a prominent feature \cite{Ryu10};
when the ray escapes from the cavity, it may either go to the infinity or collide with the other cavity so as to enter it.
For the former case we force the ray to be reflected back into the original cavity rather than allowing the ray to escape from the system.
For the latter case there exists a finite probability that the ray can also be reflected from the surface of the other cavity.
However, the ray is then forced to enter the cavity according to the rules set up in the RMDS.
These forced operations play a role of the linear map (2) in the previous section.
This shows how the RMDS works.
Although this model is simple, it notably possesses the main feature of a quasiattractor that we introduced in the previous section: 
The rays forcibly reflected back into the cavity, otherwise escaping from the cavity, indeed form quasiattracting islands.
Moreover they are shown to be directly associated with resonant modes obtained from wave calculation.

In this section, we provide how the RMDS of two coupled dielectric disks is related to the coupled standard-linear map introduced
in the previous section.
We mostly exploit the results obtained from the scattering problem of two disk hard wall billiard outside the cavity \cite{Jos92}
with Snell's law to incorporate the transmission across the cavity boundaries.

\subsection{System, selection rule, and mappings}

\begin{figure}
\begin{center}
\includegraphics[width=\figsizetwo\textwidth]{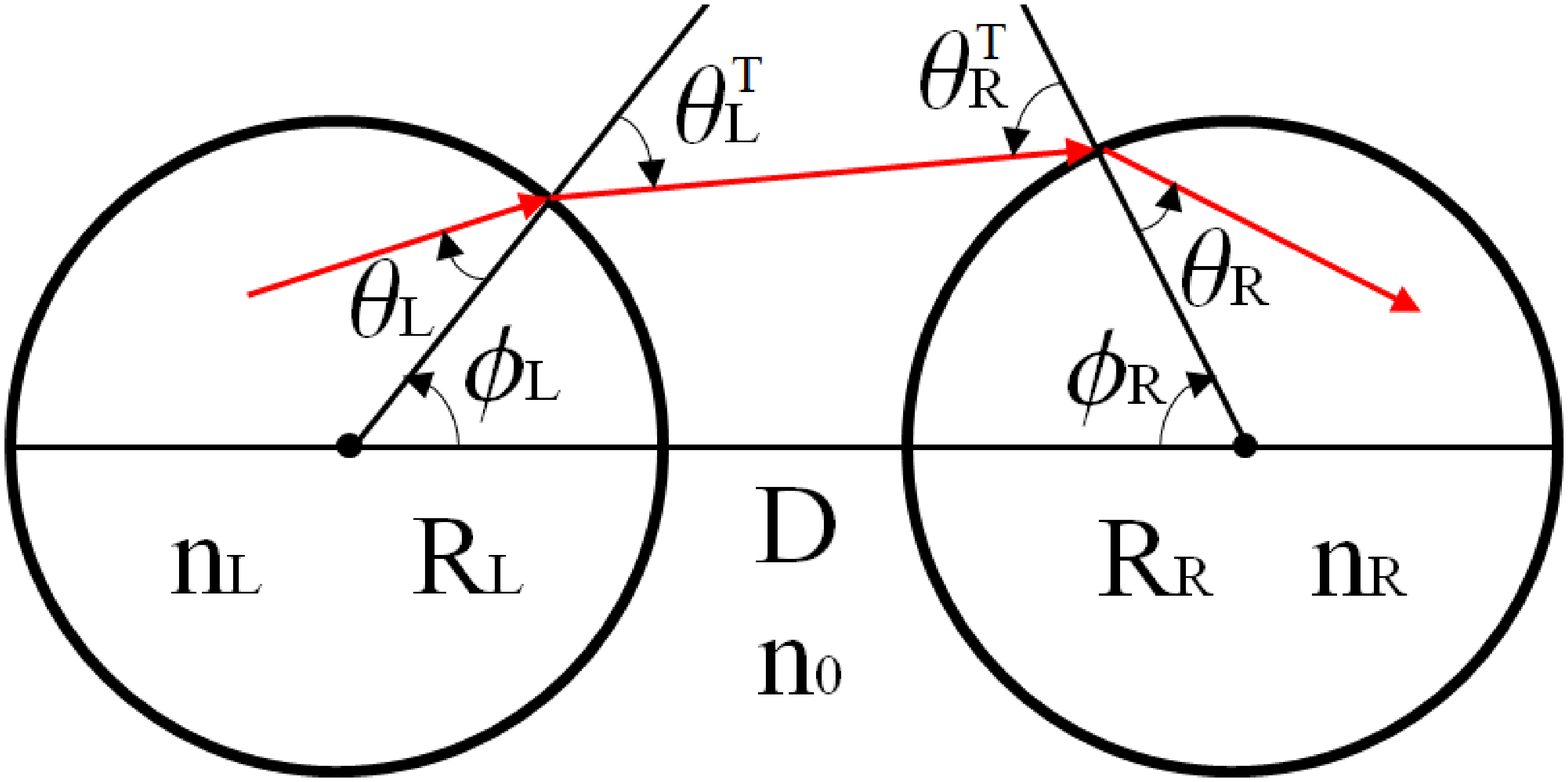}
\caption{(color online) Schematic picture of the coupled two disks.
The red arrows represent a typical trajectory of a ray transmitted from the left (L) to the right (R) disk.
Note that in the left disk $\phi_L$ and $\theta_L$ increases in the counterclockwise direction, while in the right the opposite is true.
For example, in this figure, $\phi_L$, $\phi_R$, $\theta_R$, and $\theta^{T}_R$
are positive and $\theta_L$ and $\theta^{T}_L$ are negative.
$R_L$ and $R_R$ is the radius of the left and the right disk, respectively, and $D$ is the distance between two disks.
$n_L$, $n_R$ and $n_0$ are the indices of refraction of the left disk, the right disk and the outside, respectively.}
\label{fig2}
\end{center}
\end{figure}

Two coupled disks schematically shown in Fig.~2 are described by geometric parameters, $R_R$, $R_L$, and $D$,
which denote the radius of the right and the left disk, and the inter-disk distance, respectively,
and by material parameters, namely the index of refraction of the right ($n_R$) and the left ($n_L$) disks.
For simplicity we set $n_R/n_0=n_L/n_0 \equiv n$, with the index of refraction of the medium outside the cavities, $n_0$.
The ray dynamics is traced by using Birkoff coordinate of ($\phi_{R(L)}$, $\theta_{R(L)}$), where $\phi_{R(L)}$ and
$\theta_{R(L)}$ denote the azimuthal angle and the angle of incidence in the right (left) disk, respectively.
Notice that the sign of these angles are defined differently (See Fig.~2).

The deterministic selection rule \cite{Ryu10} is given as follows.
For the sake of convenience, we assume the two disks are identical, i.e., $R \equiv R_R=R_L$
which allows us to find an analytic formulation of the problem.
We will also consider what happens when two disks are not identical below.
First of all, if the incident angle $\theta$ is larger than the critical angle $\theta_c = \arcsin (1/n) $,
rays are totally reflected from the boundary and thus cannot escape from the disk. 
Transmission of rays from one disk to the other is possible only when 
$|\theta| < \theta_c$. In addition, the escaped ray can arrive at the other disk only if $|\phi| < \pi/2$ is satisfied.
As a matter of fact, the precise condition that the ray can be transmitted from one disk to the other is obtained
by considering a simple geometry \cite{Jos92} and the Snell's law as follows 
\begin{eqnarray}
\label{con1}
&\arcsin(-1/n)<\theta<-\theta_\mathrm{min} \qquad &\textrm{for} ~ \phi_c<\phi<\pi/2, \\\nonumber
&-\theta_\mathrm{max}<\theta<-\theta_\mathrm{min} \qquad &\textrm{for} ~ -\phi_c<\phi<\phi_c, \\\nonumber
&-\theta_\mathrm{max}<\theta<\arcsin(1/n) \qquad &\textrm{for} ~ -\pi/2<\phi<-\phi_c, 
\end{eqnarray}
where
\begin{eqnarray}
\label{con2}
\phi_c=\arccos \left(\frac{2R}{2R+D}\right), \\\
\theta_\mathrm{min}=\arcsin\left(\sin({\theta_\mathrm{min}^T})/n\right), \\\
\theta_\mathrm{max}=\arcsin\left(\sin({\theta_\mathrm{max}^T})/n\right),
\end{eqnarray}
with
\begin{eqnarray}
\theta_\mathrm{min}^T=\phi&+&\arcsin\left({\frac{R\sin\phi}{l}}\right)-\arcsin\left({\frac{R}{l}}\right), \\
\theta_\mathrm{max}^T=\phi&+&\arcsin\left({\frac{R\sin\phi}{l}}\right)+\arcsin\left({\frac{R}{l}}\right),
\end{eqnarray}
where $l=\sqrt{(2R+D)^2+R^2-2R(2R+D)\cos{\phi}}$.
The area of phase space satisfying the condition (3) is marked by brown color in Fig.~3 and referred to as the outgoing area (also called the region C).
The rays located in the outgoing area of one cavity are directly mapped into the corresponding area of phase space of the other cavity,
which is called the incoming area marked by yellow color in Fig.~3.
This constitutes the heart of the selection rule playing an equivalent role of the linear map (2) in that the outgoing
and the incoming area directly correspond to the region A and B of the linear map (2), respectively.
The incoming area should form mirror image of the outgoing area due to the time-reversal symmetry and the reflection symmetry of the system. 

\begin{figure}
\begin{center}
\includegraphics[width=\figsize\textwidth]{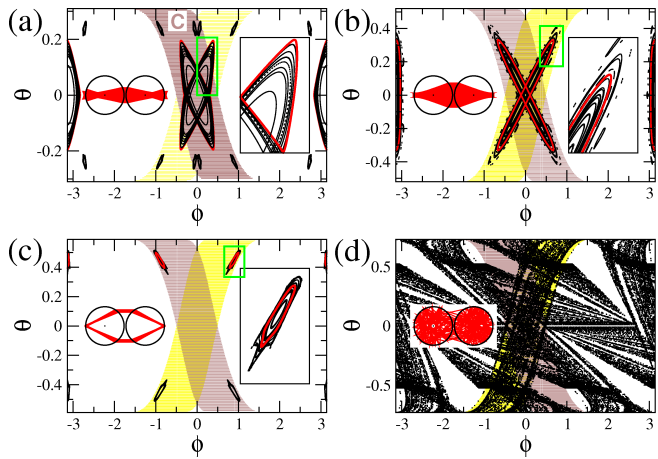}
\caption{(color online) The PSOSs of the two-disk map (TDM) for (a) $n=3.3$, (b) $n=2.0$, (c) $n=1.8$, and (d) $n=1.5$ with $D=0.1$ and $R_R=R_L=1.0$.
The right insets of (a)-(c) represent the corresponding enlarged PSOS of the part of the islands designated by the green boxes.
The left insets represent the trajectories in real space corresponding to (a)-(c) the torus denoted
as the thick red curve in the islands and (d) the chaotic quasiattractor.
The outgoing (also called the region C) and the incoming area are represented as brown (dark gray) and yellow (light gray) area, respectively.
They are symmetric over $\phi=0$ and $\theta=0$ due to the symmetric geometry shown in Fig.~2.
}
\label{fig3}
\end{center}
\end{figure}

For completeness, we provide the full details of the map. Let us consider first the case that the rays stay within one of the disks.
This is called the inside-disk dynamics, which is determined by the following simple mapping,
\begin{eqnarray}
\label{fdisk}
\phi_{n+1} &=& \phi_n + \pi-2\theta_n,\\\nonumber
\theta_{n+1} &=& \theta_n.
\end{eqnarray}

When the ray enters the outgoing area of one disk by satisfying the condition (3),
it is transmitted to reach the other disk. This transmission of the ray from one cavity to the other is described as the mapping
\begin{eqnarray}
\label{feq}
\phi_{n+1}&=&f(\phi_n,\theta_n)=T_{\phi}(\phi_n,\arcsin(n\sin(\theta_n))), \\\nonumber
\theta_{n+1}&=&g(\phi_n,\theta_n) \\\nonumber
            &=&\arcsin\left(\sin({T_{\theta}(\phi_n,\arcsin(n\sin(\theta_n)))})/n\right),
\end{eqnarray}
where $T_\phi$ and $T_\theta$ are given by \cite{Jos92}
\begin{eqnarray}
\label{feq_out}
\phi_{n+1}&=&T_{\phi}(\phi_n,\theta_n^T) \\\nonumber
          &=&\arcsin\left({\sin\phi_n+\frac{\lambda(\phi_n,\theta_n^T)}{R}\sin(\theta_n^T+\phi_n)}\right),\\\nonumber
\theta_{n+1}^T&=&T_{\theta}(\phi_n,\theta_n^T) \\\nonumber
              &=&\arcsin\left({\frac{2R+D}{R}\sin(\theta_n^T+\phi_n)-\sin(\theta_n^T)}\right),
\end{eqnarray}
with
\begin{widetext}
\begin{eqnarray}
\lambda(\phi,\theta^T)&=&(2R+D)\cos(\theta^T+\phi)-R\cos\theta^T \\\nonumber
                      & &-\sqrt{[(2R+D)\cos(\theta^T+\phi)-R\cos\theta^T]^2-(2R+D)^2+2R(2R+D)\cos\phi}.
\end{eqnarray}
\end{widetext}
The map of RMDS is thus summarized: if the condition (3) is satisfied Eq.~(\ref{feq}) is applied, otherwise Eq.~(\ref{fdisk}).
We call it two-disk map (TDM).

\begin{figure}
\begin{center}
\includegraphics[width=\figsizetwo\textwidth]{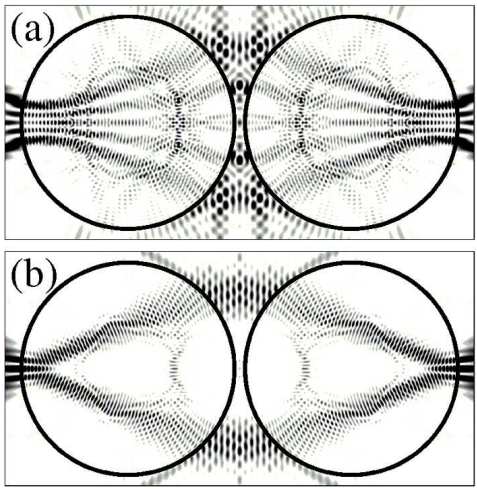}
\caption{Patterns of resonant modes obtained from wave calculation when (a) $n=2.0$ and (b) $n=1.8$,
which are directly associated with the islands of Fig.~3(b) and (c), respectively.
}
\label{fig4}
\end{center}
\end{figure}

\subsection{Quasiattractor in coupled dielectric disks}

We obtain PSOS of the TDM after some transient time for various $n$ and $R$ with $D=0.1$
using the uniformly distributed initial points $(\phi_0,\theta_0)$ in the open region, 
$-\arcsin(1/n)<\theta<\arcsin(1/n)$, of phase space.
Figure~\ref{fig3} shows the PSOSs for several $n$ with $D=0.1$ and $R_R=R_L=1.0$,
where the rays converge into the islands [Fig.~3(a)-(c)] or the fractal  structure [Fig.~3(d)] so as to form the quasiattractors.
They look similar to those found in the standard-linear map of Sec.II.
Figure~4(a) and (b) show the patterns of two resonant modes for $n=2.0$ and $n=1.8$, respectively,
which are obtained numerically using the boundary element method \cite{Wie03}.
The patterns are localized on the islands of Fig.~3(b) and (c), and closely resemble the ray trajectories presented in their insets.
It clearly shows that the resonant modes are localized on the quasiattractors.

It is noted that the PSOS obtained here, Fig.~3, is slightly different from the usual PSOS in that the points are selected
once the ray collides with the boundary irrespective of whether it is incident to or emerging from the boundary.
Usually the point is chosen only when the ray is either incident or emerging.
This explains why two islands are overlapped to form 'x' shape in Fig. 3(a). We will discuss this point below.
As $n$ decreases the islands of the quasiattractors experience bifurcation;
Firstly the island corresponding to the horizontal bouncing ball-like periodic orbit shown in the left insets of Fig.~3(a) and (b)
becomes unstable so as to be split into the hexagonal shape periodic orbit shown in the left inset of Fig.~3(c).
Secondly the hexagonal periodic orbit also loses its stability to form the fractal quasiattractor as show in Fig.~3(d).

\begin{figure}
\begin{center}
\includegraphics[width=\figsize\textwidth]{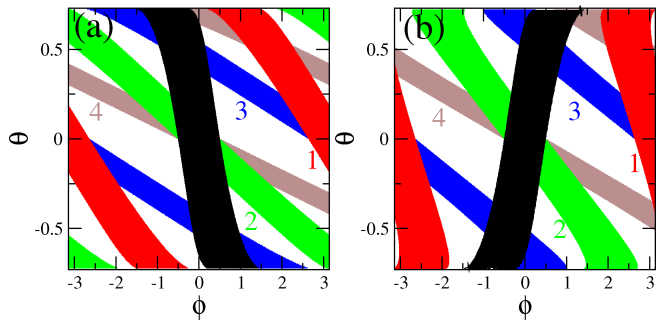}
\caption{(color online) The first (red), second (green), third (blue), and fourth (brown)
iterated sets according to Eq.~(\ref{fdisk}) starting from (a) the outgoing (black) and (b) the incoming area (black).}
\label{fig5}
\end{center}
\end{figure}

The islands in Fig.~3(a)-(c) consist of three parts, namely the parts in the outgoing ($R_o$), the incoming ($R_i$), and the reflection region ($R_r$)
in which all the ray is totally reflected in around $|\phi|=\pi$.
The rays starting from $R_r$ on the island quasiattractor is mapped to those in $R_o$, and consequently to $R_i$.
This process is repeated, that is,
$R_r \rightarrow R_o \rightarrow R_i \rightarrow R_r \rightarrow R_o \rightarrow R_i$.
If time goes backward, outgoing and incoming rays are exchanged
but the ordering rule is not affected. 
This means that the ray dynamics on the islands is reversible in time, 
in agreement with the symmetric structures of the islands.
However, in the case of a fractal-chaotic quasiattractor of Fig.~\ref{fig3} (d),
the time reversal symmetry is broken since in general there can exist two possible origins for the ray inside the cavity;
that is, one from the same cavity by reflection from the boundary and the other coming from the other cavity by refraction.
As shown in the standard-linear map of the previous section, here the fractal quasiattractor also contains empty forbidden regions related to
the short time successive iterations of outgoing and incoming regions.
Figures~\ref{fig5} (a) and (b) show the first, second, third, and fourth forward iterated sets according to Eq.~(\ref{fdisk}) from the outgoing
and incoming areas, respectively.
Iterated sets of the outgoing area in Fig.~\ref{fig5} (a) produce the white forbidden region in Fig.~\ref{fig3} (d)
whereas the iterations of the incoming area clearly leave their trace on the fractal quasiattractor.

\begin{figure}
\begin{center}
\includegraphics[width=\figsize\textwidth]{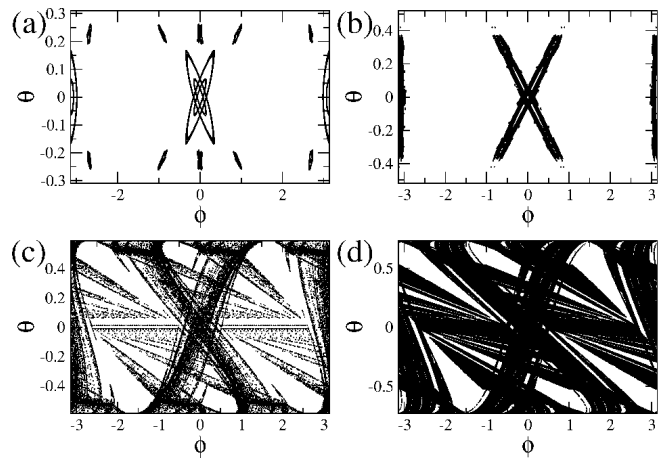}
\caption{The PSOSs of the asymmetric TDM of $R_R=1.1 R_L$ obtained in the left disk.
(a) $n=3.3$, (b) $n=2.0$, (c) $n=1.8$, and (d) $n=1.5$ with $D=0.1$.}
\label{fig6}
\end{center}
\end{figure}

So far we have considered the identical disks, i.e. $R_R = R_L$, but our results can also be applied to the case of $R_R \neq R_L$.
Figure~\ref{fig6} shows the PSOSs for various $n$ with $R_R=1.1 R_L$, in which two disks are no longer identical.
The PSOS of the left thus differs from that of the right.
In Fig.~6 the PSOS is taken from the left. $D=0.1$ is chosen as before.
The PSOS is calculated by ray tracing method considering RMDS
since the analytic expressions of TDM obtained above is only applicable for the identical disks.
For $n=1.8$ the quasiattractor of the hexagonal-shape periodic orbit is observed at $R_R=R_L$ [Fig.~3(c)],
but disappears at $R_R=1.1 R_L$ and the fractal quasiattractor takes place [Fig.~6(c)].
If $D$ varies, the topological structure of the PSOSs and the quasiattractor are also changed (not shown here).
The variations of the structures in the PSOS depending on system parameters can also be explained by the stabilities of the periodic orbits
obtained from their monodromy matrices \cite{Ryu10,Ber81,Sie90}.

\subsection{Ray splitting and generalization of the PSOS at dielectric interface}

The Birkoff coordinate of PSOS of dielectric cavities can be chosen as one of four possible components;
either the incident or the emerging ray either inside or outside cavity.
As far as a single cavity is concerned, all components are basically equal to each other since they are directly interconnected
by the law of reflection and the Snell's law \cite{Cha96,Noe97}.
However, in coupled dielectric cavities, every four components must be separately taken into account
because of more complex ray dynamics that includes refraction and subsequent reentry into another component 
as well as reflection at the dielectric boundary back into the same component.

\begin{figure}
\begin{center}
\includegraphics[width=\figsize\textwidth]{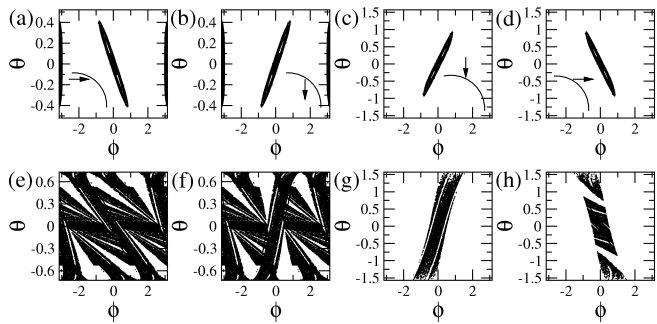}
\caption{The generalized PSOSs constructed by considering the (a, e) incident ray inside,
(b, f) emerging ray inside, (c, g) incident ray outside, and (d, h) emerging ray outside cavities,
which is schematically represented in each inset in the upper row.
We choose $n=2.0$ (the upper row) and $n=1.5$ (the lower row) with $D=0.1$ and $R_R=R_L=1.0$.
}
\label{fig7}
\end{center}
\end{figure}

We therefore introduce a generalization of the PSOS that comprises incident and emerging rays inside and outside cavities.
Figure~\ref{fig7} (a)-(d) show the generalized PSOSs of our ray model for $n=2.0$, $D=0.1$, and $R_R=R_L=1.0$.
As mentioned above, the PSOS of Fig.~3 was constructed such that the points are selected once the ray collides with the boundary.
It means that Fig.~3(b) is the combination of Fig.~7(a) and (b), which is the PSOS chosen from the incident and emerging ray inside the cavity,
respectively. Considering the nature of the coupled two disks, whose ray is either reflected from or transmitted through the boundary,
the appropriate PSOS should contain both cases. This is the reason why the PSOSs was made in this way.
The case of chaotic PSOSs with fractal quasiattracting structures is shown for $n=1.5$, $D=0.1$, and $R_R=R_L=1.0$
in Fig.~\ref{fig7} (e)-(h).
Again, Fig.~3(d) is obtained from merging Fig.~7(e) and (f).

These generalized PSOSs are also useful to study ray-wave correspondence in coupled dielectric cavities since the corresponding
quasi-eigenmode obtained from wave calculation can be represented by the generalized Husimi functions which also have
four possible realizations in the exactly same manner \cite{Hen03}.

\section{Summary}
\label{sec_IV}

We have studied a quasiattracting phenomenon in coupled dielectric cavities.
Its origin is in the ray-splitting dynamics modelled as the map based upon RMDS.
The key feature of this map can be understood by considering a simple toy model, the standard-linear map.

\section*{Acknowledgment}
We thank D.-U. Hwang, H. Kantz, S.-Y. Lee, S. Shinohara, and J. Wiersig for discussions.
We gratefully acknowledge financial support by the German Research Foundation (DFG) within the Research Unit FG 760
and the Emmy Noether Programme (M.H.).
This was supported by the NRF grant funded by the Korea government (MEST) (No.2009-0084606 and No.2010-0024644).

\section*{Appendix: Monodromy matrix of hexagonal-shaped periodic orbit}

The hexagonal-shaped periodic orbit (HSPO) is described by 12 successive mappings consisting of 6 translations (4 inside the cavity and 2 outside),
4 refractions across the boundary and 2 reflections at $\phi=\pi$.
The monodromy matrix of the HSPO can then be constructed as multiplication of the monodromy matrices of each mapping.
Note that the monodromy matrix of 2 reflections is an identity so that it can be ignored.
For $R \equiv R_R=R_L$, the incident angle of the hexagonal periodic orbit is given by $\theta=\arccos(n/2)$ when $n<2.0$.
There is no hexagonal-shaped periodic orbit if $n>2.0$.
The monodromy matrices of the translation inside and outside the cavity are
\begin{equation}
\mathbf{M_{I}}=
\left( \begin{array}{cc}
1 & -\frac{2 R}{\cos\theta} \\
0 & 1 \\
\end{array} \right)
\end{equation}
and
\begin{equation}
\mathbf{M_{O}}=
\left( \begin{array}{cc}
-1-\frac{l}{R \cos2\theta} & -\frac{l}{\cos^{2}2\theta} \\
-\frac{l}{R^2}-\frac{2\cos2\theta}{R} & -1-\frac{l}{R \cos2\theta} \\
\end{array} \right),
\end{equation}
respectively, $l=D+2R(1-\cos2\theta)$.

The monodromy matrices of the refraction from inside to outside and the opposite can be obtained from Snell's law.
\begin{eqnarray}
\mathbf{M_{B_1}}=
\left( \begin{array}{cc}
1 & 0 \\
0 & n \\
\end{array} \right)
\end{eqnarray}
and
\begin{eqnarray}
\mathbf{M_{B_2}}=
\left( \begin{array}{cc}
1 & 0 \\
0 & {{1}\over{n}} \\
\end{array} \right),
\end{eqnarray}
respectively.

Finally, the monodromy matrix of the hexagonal-shaped periodic orbit is
\begin{equation}
\mathbf{M}=\mathbf{M_I}\mathbf{M_{B_2}}\mathbf{M_O}\mathbf{M_{B_1}}\mathbf{M_{I}}\mathbf{M_{I}}
\mathbf{M_{B_2}}\mathbf{M_O}\mathbf{M_{B_1}}\mathbf{M_I}.
\end{equation}
The stability of the hexagonal-shaped periodic orbit depends on the value of $\mathrm{Tr}~\mathbf{M}$.
For $R=1.0$ and $D=0.1$, in the case of $\mathrm{Tr}~\mathbf{M} > 2$ when $n \lesssim 1.688$, the periodic orbit is linearly unstable
but in the case of $\mathrm{Tr}~\mathbf{M} < 2$ when $n \gtrsim 1.688$, the periodic orbit is linearly stable.


\begin{thebibliography}{150}
\bibitem{remark_A-is-minimal} For mathematical completeness, we give the third property:
$A$ is minimal, i.e., there is no proper subset of $A$ that satisfies first and second conditions.
\end{thebibliography}
\end{document}